\documentclass[usenatbib]{mn2e}
\usepackage{natbibmnfix,graphicx,times}

\usepackage{epstopdf}
\usepackage{natbib}
\newcommand{\HI}{\mathrm{H\,I}}
\newcommand{\HeI}{\mathrm{He\,I}}
\newcommand{\HeII}{\mathrm{He\,II}}
\newcommand{\HeIII}{\mathrm{He\,III}}
\newcommand{\HIa}{H\,{\sevensize{\textbf{I}}}\,\,}

\newcommand{\HeIa}{He\,\sevensize{\textbf{I}}\normalsize\,\,}
\newcommand{\HeIIa}{He\,{\sevensize{\textbf{II}}}\,\,}
\newcommand{\HeIIb}{He\,\sevensize{\textbf{II}}\large\,\,}
\newcommand{\HeIIc}{He\,\sevensize{\textbf{II}}\large}
\newcommand{\HeIIIa}{He\,\sevensize{\textbf{III}}\normalsize\,\,}
\newcommand{\GHI}{\Gamma_{\HI}}
\newcommand{\GHeII}{\Gamma_{\HeII}}
\newcommand{\etath}{\eta_{\mathrm{thin}}}

\newcommand{\lya}{Ly$\alpha$ }

\newcommand{\apjl}{ApJ}

\newcommand{\mnras}{MNRAS}

\title[Effect of Fluctuations on the Helium-Ionizing Background]{The Effect of Fluctuations on the Helium-Ionizing Background} 
\author[F.~B. Davies, S.~R. Furlanetto]{Frederick B. Davies\thanks{davies@astro.ucla.edu}, Steven R. Furlanetto\thanks{sfurlane@astro.ucla.edu}\\
Department of Physics \& Astronomy, University of California, Los Angeles, Box 951547, Los Angeles, CA 90095}

\begin{document}

\maketitle

\begin{abstract}
Interpretation of \HeIIb Ly$\alpha$ absorption spectra after the epoch of \HeIIb reionization requires knowledge of the \HeIIb ionizing background. While past work has modelled the evolution of the average background, the standard cosmological radiative transfer technique assumes a uniform radiation field despite the discrete nature of the (rare) bright quasars that dominate the background. We implement a cosmological radiative transfer model that includes the most recent constraints on the ionizing spectra and luminosity function of quasars and the distribution of IGM absorbers. We also estimate, for the first time, the effects of fluctuations on the evolving continuum opacity in two ways: by incorporating the complete distribution of ionizing background amplitudes into the standard approach, and by explicitly treating the quasars as discrete -- but isolated -- sources. Our model results in a \HeIIb ionization rate that evolves steeply with redshift, increasing by a factor $\sim2$ from $z=3.0$ to $z=2.5$. This causes rapid evolution in the mean \HeIIb Ly$\alpha$ optical depth -- as recently observed -- without appealing to the reionization of \HeIIc. The observed behaviour could instead result from rapid evolution in the mean free path of ionizing photons as the helium in higher \HIa column density absorbers becomes fully ionized.
\end{abstract}


\section{Introduction}

The ionizing background is crucial for understanding many aspects of large-scale structure and galaxy formation at high redshifts. For example, unraveling the physical density structure of the \lya forest (which contains most of the the intergalactic medium, or IGM, at $z \ga 2$) requires knowledge of the ionization state of the intervening material \citep{Rauch1998,Meiksin2009}. It is also crucial for understanding the abundance and distribution of heavy elements in the IGM, whose ionization states depend sensitively on the local metagalactic radiation field (e.g., \citealt{Songaila1998,Songaila2005,KimMetal2002,Aguirre2004,Bolton2011}). Additionally, the ionizing background is an important input parameter for cosmological simulations because it regulates the dominant heating and cooling in the IGM \citep{Dave1999,Springel2003}, which forms the fuel supply for later galaxy formation. Finally, the ionizing background holds important clues about galaxies and quasars, because they are the dominant sources behind it. Precise measurements can constrain the star formation rate, the escape fraction of ionizing photons from galaxies, and the importance of luminous quasars \citep{Madau1999,FG2008b,FG2009,HM2012}.

Perhaps most importantly, the ionizing background is tied inextricably to the reionization process, when the global ionization state of intergalactic atoms changes rapidly. For example, measurements of the \HIa ionizing background at $z \sim 5$--$6$ show that hydrogen reionization appears to proceed relatively slowly \citep{BH2007}. Its properties will also be crucial for understanding \HeIIa reionization, which is due to bright quasars \citep{Sokasian2003, FP2008, McQuinn2009}. Based on studies of the effective optical depth of the \HeIIa \lya forest, the reionization of \HeIIa in the universe seems to have completed at $z\sim3$ \citep{Reimers1997,Kriss2001,Zheng2004,Shull2004}. The evolution of the ionizing background during and after \HeIIa reionization is critical to interpreting new and upcoming \HeIIa \lya forest results from \textit{HST/COS} \citep{Shull2010,Worseck2011,Syphers2012}. Theoretical calculations have attempted to address this evolution by semi-analytic modelling \citep{DF2009,FD2010} and hydrodynamic simulations of the IGM \citep{Sokasian2003,Bolton2006,Paschos2007,McQuinn2009}.

There is a long history of calculations to estimate the properties of the metagalactic ionizing radiation field. \citet{HM1996} made a landmark study of the ionizing background using a cosmological radiative transfer model for ionizing photons traveling through a clumpy IGM. By combining state-of-the-art constraints on the distribution of ionizing sources and the absorber distribution of the IGM, \citet{HM1996} were able to compute the evolving ionizing background of \HIa and He\,{\sevensize{\textbf{II}}}. Further studies \citep{Fardal1998,FG2009,HM2012} have updated this framework with new constraints on the population of ionizing sources and the distribution and properties of IGM absorbers. However, all of these studies treated the ionizing background (and its sources and sinks) as \emph{uniform} components, which is a reasonable approximation for the \HIa background (at least at low and moderate redshifts; \citealt{MW2004}) but is a poor approximation when bright, rare sources dominate the emissivity (as is the case for quasars and the \HeIIa ionizing background).

\citet{Fardal1998} showed how the relatively large mean separation of \HeIIa ionizing sources could contribute to the significant observed fluctuations in the ionizing background and hence in the observable \HeIIa \lya effective optical depth.   An analytic description of variations in the metagalactic radiation field was introduced by \citet{Zuo1992}, expanded by \citet{MW2003}, and later used by \citet{Furlanetto2009} to study fluctuations in the \HeIIa ionizing background. Despite this theoretical interest, there has been no effort to include the effect of these fluctuations on the ionizing continuum opacity within a cosmological radiative transfer model. In this work, we attempt to show the self-consistent effect of these fluctuations on the mean ionizing background.

We begin in Section 2 with a description of our implementation of a cosmological radiative transfer model to calculate self-consistently the \HeIIa ionization rate. Then, in Section 3, we present the results of our model. In Section 4, we use the results from that model to calculate the evolution of the \HeIIa effective optical depth and compare it to observations. We discuss our model assumptions and compare to previous work in Section 5. We conclude in Section 6.

In our calculations, we assume the following cosmology: $\Omega_m = 0.26$, $\Omega_\Lambda = 0.74$, $\Omega_b = 0.044$, and $h = 0.74$ \citep{Dunkley2009}. All distances are given in comoving units unless otherwise specified.

\section{Inputs/Methods}
\subsection{Cosmological Radiative Transfer} \label{sect:method}

To calculate the \HeIIa ionizing background, we employ a cosmological radiative transfer model \citep{HM1996}.  By considering photon conservation in a comoving volume element, the specific intensity of ionizing radiation $J_\nu$ behaves as 
\begin{equation}\label{eqn:rad}
\left(\frac{\partial}{\partial t} - \nu H \frac{\partial}{\partial \nu}\right) J_\nu = -3HJ_\nu - c \alpha_\nu J_\nu + \frac{c}{4\pi}\epsilon_\nu ,
\end{equation}
where $H(t)$ is the Hubble parameter, $c$ is the speed of light, $\alpha_\nu$ is the absorption coefficient (with $d \tau_\nu = \alpha_\nu dl$ and $dl$ the proper line element), and $\epsilon_\nu$ is the proper emissivity. This approach assumes that each volume element can be described as an isotropic source and sink of radiation through $\epsilon_\nu$ and $\alpha_\nu$, respectively: we will revisit this assumption later on. 
The solution to the cosmological radiative transfer equation is 
\begin{equation}\label{eqn:jnu}
J_{\nu_0}(z_0) = \frac{1}{4\pi}\int_{z_0}^{\infty} dz \frac{dl}{dz}\frac{(1+z_0)^3}{(1+z)^3}\epsilon_\nu(z)\exp[-\bar{\tau}(\nu_0,z_0,z)].
\end{equation}
where $dl/dz = c/((1+z)H(z))$ is the proper line element, $\nu = \nu_0 (1+z)/(1+z_0)$, and $\bar{\tau}$ is the effective optical depth experienced by a photon at frequency $\nu_0$ and redshift $z_0$ since its emission at redshift $z$.  $\bar{\tau}$ is calculated using $e^{-\bar{\tau}} = \langle e^{-\tau} \rangle$ averaging over all lines of sight.  For Poisson-distributed absorbers with \HIa column density $N_\HI$ this opacity is \citep{Paresce1980} 
\begin{equation}\label{eqn:tbar}
\bar{\tau}(\nu_0,z_0,z) = \int_{z_0}^{z} dz' \int_{0}^{\infty} dN_{\HI} {\partial^2N\over\partial N_{\HI}\partial z'} (1-e^{-\tau_\nu}),
\end{equation}
where ${\partial^2N/\partial N_{\HI}\partial z} \equiv f(N_\HI,z)$ is the column density distribution function (CDDF) of neutral hydrogen absorbers. The most common simple form of the CDDF is a power law in column density and redshift: $f(N_\HI,z)\propto N_\HI^{-\beta}(1+z)^\gamma$, but we will allow more sophisticated models as well (see \S~\ref{sect:CDDF}).

The optical depth of an absorber to ionizing photons of frequency $\nu$ is given by
\begin{equation}\label{eqn:taunu}
\tau_\nu = N_{\HI}\sigma_\HI(\nu) + N_{\HeI}\sigma_\HeI(\nu) + N_{\HeII}\sigma_\HeII(\nu) ,
\end{equation}
where $N_i$ are the column densities and $\sigma_i$ are the photoionization cross-sections of ion $i$. Because only the column density distribution of $N_\HI$ has been measured, we use a model for the relationship between $N_\HI$ and $N_\HeII$ to calculate the \HeIIa ionizing opacity (see \S~\ref{sect:absmodel}). In the frequency range contributing to the \HeIIa ionizing background ($\nu > \nu_\HeII = 4\,\nu_\HI$) we assume the contribution to the optical depth from \HeIa is negligible following \citet{FG2009}. Finally, the ionization rate for \HeIIa is given by
\begin{equation}\label{eqn:Gamma}
\Gamma_\HeII(z) = 4\pi\int_{\nu_\HeII}^{\infty}\frac{J_\nu(z)}{h\nu}\sigma_\HeII(\nu) d\nu ,
\end{equation}
where $\nu_\HeII$ is the ionization threshold of He\,{\sevensize{\textbf{II}}}.

In our model, we do not explicitly calculate the \HIa ionization rate, as that calculation depends strongly on poorly constrained models of the escape fraction of ionizing photons from star-forming galaxies (see e.g. \citealt{HM2012}). Because the detailed evolution of $\GHI$ is not the focus of this work, we instead adopt an empirical estimate of the \HIa ionization rate from measurements of the \lya forest \citep{FG2008b}, which appears to be fairly constant over our redshift range of interest ($z\sim2$--$4$). We have ensured that our fiducial value for $\GHI$ is consistent with our fiducial quasar emissivity and CDDF; that is, the value of $\GHI$ calculated in our fiducial model with quasars only is less than the value we assume in the fiducial $\GHeII$ calculation.

Because the ionizing background and the continuum opacity are interrelated through the conversion of $N_\HI$ to $N_\HeII$ described in the next section, the procedure must be iterated over the entire redshift range until convergence. The result of this cosmological radiative transfer model as presented in this section will be referred to as the "uniform" background model in the rest of the paper.

\subsubsection{Absorber Ionization Structure: N$_\HI$ to N$_\HeII$} \label{sect:absmodel}

The relationship between $N_\HI$ and $N_\HeII$ is usually parameterized by the quantity $\eta=N_\HeII/N_\HI$ \citep{M-E1993}. In the optically thin case, $\eta$ is given by 
\begin{equation}\label{eqn:etath}
\etath = \frac{\GHI}{\GHeII}\frac{\alpha_\HeII^A}{\alpha_\HI^A}\frac{Y}{4X} ,
\end{equation}
where $\alpha_\HI^A$ and $\alpha_\HeII^A$ are the case-A recombination coefficients of \HIa and \HeIIa, and $X=0.75$ and $Y=0.25$ are the hydrogen and helium mass fractions, respectively. In an optically thin environment, photons produced by recombinations to the ground state of \HeIIa will escape from the local medium, hence our choice of case-A recombination coefficients. Note, however, that these coefficients enter only in the ratio, so this choice does not have any significant effect.

To more generally translate \HIa column densities into \HeIIa, we adopt a fit to numerical simulations that accounts for self-shielding in neutral hydrogen systems \citep{Fardal1998, FG2009},
\begin{equation}\label{eqn:HtoHe}
\frac{Y}{16X}\frac{\tau_\HI}{1+A\tau_\HI} I_\HI = \tau_\HeII + \frac{\tau_\HeII}{1+B\tau_\HeII} I_\HeII,
\end{equation}
where $\tau_i = \sigma_i N_i$, $A = 0.15$ and $B = 0.2$ are fitting coefficients used by \citet{FG2009}, and $I_i = \Gamma_i/n_e\alpha_i^A$ with $n_e = 1.4 \times 10^{-3}$ cm$^{-3} (N_\HI/10^{17.2}$ cm$^{-2})^{2/3}(\Gamma_\HI/10^{-12}$ s$^{-1})^{2/3}$ \citep{Schaye2001}.  At small \HIa column densities ($N_\HI \la 10^{15}$ cm$^{-2}$), $N_\HeII = \etath N_\HI$ as expected. \HeIIa becomes optically thick to ionizing radiation for larger column densities ($N_\HI \sim 10^{15}$--$10^{17}$ cm$^{-2}$), so $\eta$ increases by a factor of a few as more \HeIIa forms while hydrogen remains highly ionized.  Then, for $N_\HI \ga 10^{17}$ cm$^{-2}$, $\eta$ steeply drops as the systems become optically thick to \HIa ionizing photons.

For systems with $N_\HI > 10^{18}$ cm$^{-2}$, the numerical fit systematically under-predicts the amount of \HeIIa from the original model (see Figure 1 of \citealt{FG2009}). For frequencies near $\nu_\HeII$, the opacity is unaffected because these high $N_\HI$ systems are still optically thick due to \HIa absorption. However, for $\nu \ga 2.5\,\nu_\HeII$, absorbers with $N_\HI \sim 10^{19}$--$10^{20}$ cm$^{-2}$ start to become optically thin due to their relative lack of \HeIIa. Fortunately, the total ionization rate only changes slightly because the range of affected column densities is small and the vast majority of ionizations occur at lower frequencies ($\sigma_\HeII \propto \nu^{-3}$).

\citet{HM2012} applied a similar method to fit the absorber structure that considers the average $\Gamma$ within absorbers instead of the external ``optically-thin" $\Gamma$. While their method provides a better fit to the numerical models at $N_\HI > 10^{18}$ cm$^{-2}$, it differs from the \citet{FG2009} model only in the details for the more important $\tau_\HeII\sim1$ ($N_\HI \sim 10^{16}$ cm$^{-2}$) absorbers. This is an example of one of the systematic uncertainties in our procedure: these models for $\eta$ must assume physical characteristics for the absorbers (densities, temperatures, and geometry, for example) that are both uncertain and simplifications of the true IGM physics.  For concreteness, the numerical absorber model from \citet{FG2009} assumes uniform density semi-infinite slabs with a thickness determined by the local Jeans length (at $T=20,000$~K) in photoionization equilibrium with both an external radiation background and internal recombination processes.

\subsubsection{Recombination Emissivity} \label{sect:recombs}

Recombinations of \HeIIIa to the ground state of \HeIIa will produce ionizing continuum radiation. Although the recombination rate in a uniform density medium can easily be estimated from ionization equilibrium, the real universe requires a more detailed approach for two reasons. First, density inhomogeneities in the IGM substantially boost the recombination rate.  We can model this by integrating over the \HIa column density distribution of the Ly$\alpha$ forest. Second, recombination photons produced inside optically thick absorbers will not escape to affect the IGM. 

We model the recombination emissivity of IGM absorbers with a numerical fit to the radiative transfer models of \citet{FG2009}. The emergent specific intensity from an absorber with \HeIIa column density $N_\HeII$ can be approximated by
\begin{eqnarray}
I^\mathrm{rec}_\nu(N_\HeII) &=& \frac{h\nu}{4\pi}\left(1-\frac{\alpha^B_\HeII}{\alpha^A_\HeII}\right)\Gamma_\HeII \phi_{\nu,\mathrm{rec}} \nonumber \\
&& \times\, N_T\left(1-e^{N_\HeII/N_T}\right) ,
\label{eq:irec}
\end{eqnarray}
where the second factor is the fraction of ionizations to the ground state and the local ionization rate is $\Gamma_\HeII$. $N_T = 10^{17.3}$ cm$^{-2}$ is the approximate threshold \HeIIa column density above which the emission becomes saturated by absorption within the absorber itself (the decline at larger columns is approximated by the last factor). $\phi_{\nu,\rm{rec}}$ is the normalized recombination emission profile:
\begin{equation}
\phi_{\nu,\rm{rec}} \propto \nu^{-1} e^{-h\nu/k_BT} \theta(\nu-\nu_\HeII),
\end{equation}
where $\theta(x)$ is the Heaviside step function.  The effective frequency width of this emission is $\Delta\nu/\nu \sim k_BT/h\nu_\HeII \sim 0.03$, which limits the distance these photons can travel to $\la 30$ Mpc before redshifting below the \HeIIa ionizing edge.

The total proper emissivity from recombinations is then
\begin{eqnarray}
\epsilon_{\nu,\rm{rec}}(z) = \frac{dz}{dl} \int_0^\infty dN_\HI\, f(N_\HI,z)\, 4\pi I^\mathrm{rec}_\nu(N_\HeII),
\end{eqnarray}
where the intensity depends implicitly on $N_\HI$ through the conversion factor $\eta$. We include the recombination emissivity in the cosmological radiative transfer calculation by simply adding it to the emissivity from quasars, ignoring the difference in spatial distribution.

We note here that the recombination photons can have a much larger effect on the ionizing background than one might naively expect from their emissivity. As we shall see later, increasing the emissivity also increases the mean free path of ionizing photons, which amplifies the effect of the additional ionizing photons.  We will explore this issue further in \S~\ref{sect:recresults}.

\subsection{Mean Free Path}\label{sect:mfpmethod}

The opacity per unit redshift, $d\bar{\tau}/dz$, was integrated in equation~(\ref{eqn:tbar}) to calculate the total opacity between two redshifts:
\begin{equation}\label{eqn:dtdz}
	 	{{d\bar{\tau}}\over dz} = \int_{0}^{\infty} dN_{\HI}  f(N_\HI,z) (1-e^{-\tau_\nu(\GHeII)}),
\end{equation}
where the absorber opacity as a function of $N_\HI$, $\tau_\nu$, depends on $\GHeII$ through the absorber model in Section~\ref{sect:absmodel}. At a given redshift, $d\bar{\tau}/dz$ describes the local opacity due to the forest of individual absorbers in the IGM. By inverting this quantity and converting from redshift difference to a comoving distance, we find the distance per unit optical depth, which is simply the mean free path:
\begin{equation}\label{eqn:mfp}
     	\lambda_{\mathrm{mfp}}(\nu,z) = {dl\over dz}\left({{d\bar{\tau}}\over dz}\right)^{-1}.
\end{equation}
If $f(N_i,z) = N_0N_i^{-\beta}(1+z)^{\gamma}$ and $\sigma_i = \sigma_0 (\nu/\nu_i)^{-3}$, the comoving mean free path reduces to
\begin{eqnarray}\label{eqn:lambda}
\lambda_\mathrm{mfp}(\nu,z) &\approx& \frac{(\beta-1)c}{\Gamma_G(2-\beta)N_0\sigma_0^{\beta-1}}\left(\frac{\nu}{\nu_i}\right)^{3(\beta-1)} \nonumber \\
&& \times\frac{1}{(1+z)^{\gamma}H(z)},
\end{eqnarray}
where $\Gamma_G$ is the Gamma function. The redshift dependence of the mean free path in this simplified model is then $\lambda_\mathrm{mfp} \propto (1+z)^{-(\gamma+1.5)}$. This power law dependence is a good approximation to describe the evolution of the mean free path of \HIa ionizing photons ($\lambda_\HI$) in our model because the \HIa CDDF is fixed, but we find that it fails to capture the more complicated $\GHeII$-dependent evolution of the mean free path of \HeIIa ionizing photons ($\lambda_\HeII$; see \S~\ref{sect:mfpresults}).

Recent efforts by \citet{Prochaska2009} and \citet{OM2012} have directly measured the \HIa ionizing mean free path of the IGM near $z\sim4$ and $z\sim2$ respectively. For an identical distribution of absorbers, they would report different values than obtained by our approach because they define the mean free path as the distance traveled by a photon through the \emph{evolving} IGM while it redshifts with the cosmic expansion, rather than the path that could be traveled if the IGM and photon retained their original properties (as is the usual definition for theoretical work). We follow the latter definition here.

\subsection{Fluctuations} \label{sect:fluct}

In a smooth, fully-ionized IGM, the intensity of ionizing radiation from an individual quasar falls as $\exp[-r/\lambda_\mathrm{mfp}]/r^2$. Given a distribution of quasar luminosities and a mean free path, a probability distribution of intensities can be computed assuming random placement of quasars following Poisson statistics \citep{Zuo1992,MW2003}. The effects of this distribution on the \emph{mean} ionizing background have not previously been considered. The next stage in our model is therefore to incorporate the distribution (in a somewhat ad hoc manner) in order to understand better the implications of this fluctuating background.

We use the \citet{Hopkins2007} $B$-band quasar luminosity function (QLF) to describe the distribution of relative quasar luminosities, assuming an average quasar spectral energy distribution such that the specific luminosity at the \HIa ionizing edge is proportional to the $B$-band specific luminosity ($L_B$), then extrapolating to the \HeIIa ionizing edge by a spectral index $\alpha$.  Additionally, while the effect is relatively minor \citep{Furlanetto2009}, we convolve the quasar luminosity function with a distribution of far-ultraviolet spectral indices that roughly matches observations by \citet{Telfer2002}: a Gaussian distribution over $0.5 < \alpha < 3.5$ with central value $\bar{\alpha} = 1.5$ and $\sigma_\alpha = 0.7$. Note that the asymmetric bounds on $\alpha$ lead to an average spectral index of $\alpha\simeq1.6$ consistent with our fiducial value (described later in \S~\ref{sect:emissparam}). In detail, the average ratio between the emissivity at 1 and 4 Ryd will be somewhat higher than the ratio for a $\alpha=1.6$ spectrum, but we fold this uncertainty into the ionizing background normalization uncertainty described in \S~\ref{sect:results}.

We use the method of characteristic functions from \citet{MW2003} to determine the probability distribution of intensity, $f(J)$, then scale linearly to $\Gamma$ by $\Gamma = J \times \langle \Gamma \rangle / \langle J \rangle$ \citep{Furlanetto2009}.  The last assumption of proportionality between the intensity of radiation and the ionization rate is not strictly true; the intensity at higher frequencies should be more uniform because the mean free path is much larger, although the effect is modest in practice \citep{Dixon2012}.  In our calculation of $f(\Gamma)$ we use the mean free path of the ``average" \HeIIa ionizing photon, $\bar{\lambda}_\mathrm{HeII} = \lambda_\mathrm{mfp}(\bar{\nu})$, where $\bar{\nu}$ is defined by
\begin{equation}\label{eqn:nubar}
\bar{\nu}\Gamma_\HeII = 4\pi\int_{\nu_\HeII}^{\infty}\nu\times\frac{J_\nu(z)}{h\nu}
\sigma_\HeII(\nu) d\nu,
\end{equation}
in an attempt to average over the frequency dependence of the background fluctuations. In general, $\bar{\lambda}_\HeII$ is substantially larger than $\lambda_\HeII$, so our approach provides a conservative estimate when used to calculate the amplitude of ionizing background fluctuations.  

\begin{figure}
\begin{center}
\resizebox{8cm}{!}{\includegraphics{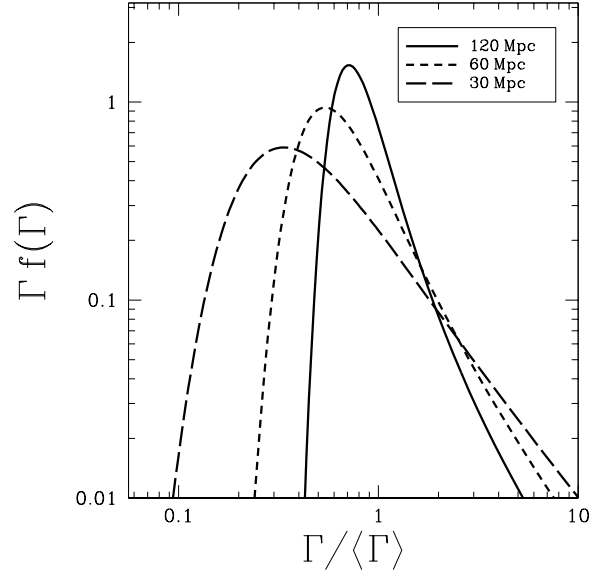}}\\
\end{center}
\caption{Distribution of ionization rates at $z=3.0$ for $\lambda_\mathrm{mfp}=30,60,120$ Mpc (long-dashed, dashed, and solid, respectively).}
\label{fig:fgamma}
\end{figure}

Figure~\ref{fig:fgamma} shows how the $f(\Gamma)$ distribution varies with mean free path. When the mean free path decreases, the peak of the distribution skews towards smaller $\Gamma$ relative to the mean. For $\Gamma$ below the mean, we find that the \HeIIa opacity of each absorber will increase, with the total opacity increasing as, roughly, $d\bar{\tau}/dz\propto\Gamma^{-2/3}$ using the \HeIIa absorber model of Section~\ref{sect:absmodel}. Because this relationship between the ionization rate and opacity is more gentle than linear, the skewness of the $f(\Gamma)$ distribution results in an average opacity that is \emph{higher} than the opacity at the mean $\Gamma$. That is, the lower opacity in rare high $\Gamma$ regions does not counteract the higher opacity in common low $\Gamma$ regions. We explore this effect in the following discussion.

We incorporate these fluctuations into our ionizing background model by averaging the opacity $d\bar{\tau}/dz$ 
(equations~\ref{eqn:tbar}, \ref{eqn:dtdz}) over the distribution $f(\Gamma)$:
\begin{equation}\label{eqn:fluct}
\left\langle\frac{d\bar{\tau}}{dz}\right\rangle = \int_0^\infty 
\frac{d\bar{\tau}}{dz}(\Gamma) f(\Gamma) d\Gamma ,
\end{equation}
where $d\bar{\tau}/dz$ depends on $\Gamma$ through the absorber prescription in Section~\ref{sect:absmodel} and $f(\Gamma)$ is initialized with the mean free path calculated in the uniform model. This process is repeated using the same $f(\Gamma)$ for each frequency in equations~(\ref{eqn:jnu}) and~(\ref{eqn:Gamma}) to modify the ionizing continuum opacity at each redshift. The fractional increase in the opacity due to the integral over $f(\Gamma)$ is larger for smaller $\lambda_\mathrm{mfp}$, reaching $\sim40\%$ in our fiducial model if $\lambda_\mathrm{mfp}$ is equal to the average distance between bright sources at $z=3$ ($\sim45$ Mpc; see \S\ref{sect:fluctresults}). Because the modified opacity leads to new values for $\GHeII(z)$ and $\lambda_\mathrm{mfp}(z)$, we iterate this process using the new $\lambda_\mathrm{mfp}(z)$ to generate $f(\Gamma)$ and using the new $\GHeII(z)$ to calculate $d\bar{\tau}/dz(\Gamma,z)$.

Unfortunately, as presented above, the $\GHeII$ calculation does not converge to a non-zero value; the added opacity from the $f(\Gamma)$ prescription causes the iterative procedure to drive $\GHeII$ down to zero. At relatively high redshifts ($z\ga3.5$) the mean free path is short enough ($\lambda_\mathrm{mfp} \la 50$ Mpc) that integrating over $f(\Gamma)$ greatly increases the opacity. In practice, this increased opacity at high redshift propagates small values of $\GHeII$ to lower redshifts, and the iterative effect pulls $\Gamma$ down to zero at \emph{all} redshifts. Even when the ionizing background is calculated assuming local emission and absorption of photons (i.e. without an integral over redshift as in equation~\ref{eqn:jnu}) via the absorption-limited approximation $J_\nu(z) = \epsilon_\nu(z)\lambda_\mathrm{mfp}(\nu,z)/(4\pi)$ \citep{MW2003}, the divergence to zero remains at $z\ga3.2$.

The reason our procedure breaks down is actually obvious: our cosmological radiative transfer model assumes that ionizing photons are emitted uniformly throughout the universe (with a constant $\epsilon_\nu$ in equation~\ref{eqn:rad}), but the real quasar sources are of course point-like. Since the ionizing background near a source is much stronger than the average, the local IGM will be less opaque to ionizing photons, and the quasar photons will penetrate farther into the IGM -- increasing the ionizing background. Additionally, our model assumes that the path traversed by an ionizing photon fully samples the distribution of ionization rates given by $f(\Gamma)$, but within a quasar proximity region this is not accurate, as the radiation profile is smoothly decreasing. To quantify the minimum effect these transparent proximity regions must have on the mean background, we consider a simple model where the ionizing background is calculated as the sum of isolated source ionization rate profiles.\footnote{For simplicity, we will ignore the finite lifetimes of quasars in our calculation. In reality, these finite lifetimes limit the extent of an individual quasar's radiation field. However, the radiation field will continue to propagate outward even after the quasar shuts off, following the profile that we describe here. The statistical results we describe here are therefore unaffected by a finite lifetime.}

\subsection{Minimum Background Model} \label{sect:mingamma}

\begin{figure}
\begin{center}
\resizebox{8cm}{!}{\includegraphics{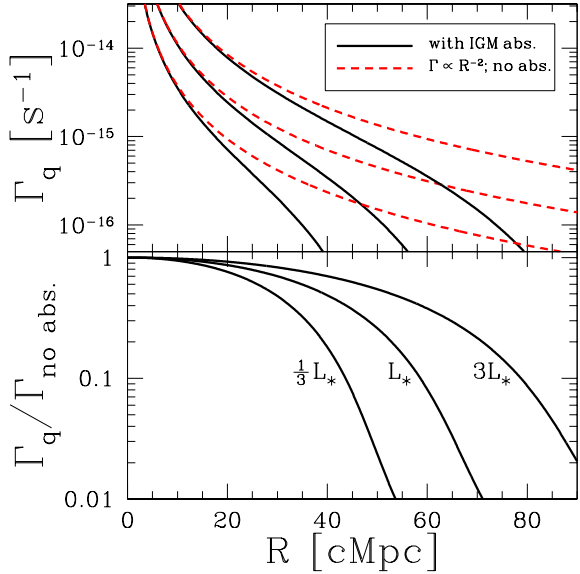}}\\
\end{center}
\caption{Line-of-sight equilibrium ionization rate profile for $L =  1/3$~$L_\ast,L_\ast,3L_\ast$ (from bottom to top) quasars at $z=3$ with IGM continuum absorption (solid black) and without (dashed red). In all cases, the quasars are assumed to be isolated (i.e., with no contribution from a metagalactic background).}
\label{fig:gammaprofile}
\end{figure}

In the absence of an external ionizing background and ignoring the cosmological redshift of ionizing photons, the equilibrium ionization rate profile along a sightline from a single quasar, $\Gamma_q(R)$, is given by
\begin{equation}
\Gamma_q(R) = \int_{\nu_\HeII}^\infty \frac{L_\nu}{4\pi R^2 h\nu}\sigma_\nu \exp[-\tau_\nu(R)]d\nu,
\end{equation}
where $\tau_\nu(R)$ is the optical depth at frequency $\nu$ from the IGM at $r < R$,
\begin{equation} \label{eqn:taunur}
\tau_\nu(R) = \int_0^R \frac{d\tau}{dz}(\nu,\Gamma_q(r))(\frac{dl}{dz})^{-1} dr,
\end{equation}
and assuming $L_\nu \propto \nu^{-1.6}$ as the mean quasar spectrum for simplicity.

Figure~\ref{fig:gammaprofile} shows the ionization rate profiles for $L =  1/3$~$L_\ast,L_\ast,3L_\ast$ quasars (from bottom to top) at $z = 3$. At small radii, the effective mean free path is very large, so $\Gamma_q \sim R^{-2}$.  However, once $\Gamma_q$ is small enough such that $R \sim \lambda_\mathrm{mfp}(\Gamma_q)$, the ionization rate drops sharply. Thus, each quasar has a characteristic radius beyond which it generates very few ionizations, effectively a recombination-limited ``proximity zone."

This ionization rate profile, integrated from small to large radius, can be calculated without detailed radiative transfer because all of the ionization state and absorption properties are contained in our prescription for the clumpy IGM through the CDDF and absorber structure from Section~\ref{sect:absmodel}. In a physical sightline, the attenuation in the IGM will be dominated by random encounters with absorbing clouds, so a more accurate description could be obtained by radiative transfer through a realistic IGM density field. We assume ionization equilibrium in the average IGM for simplicity.

The sum of these isolated quasar profiles should provide a \emph{minimal} estimate of the mean background consistent with the CDDF and the QLF, because they ignore collective effects from the overlap of the proximity zones. Armed with the $\Gamma_q$ profiles as a function of quasar luminosity, we can calculate this minimum mean background by integrating over the QLF and averaging over position,
\begin{equation}
\Gamma_\mathrm{min} = \int_0^\infty \left(\int_{L_\mathrm{min}}^\infty \Gamma_q(R,L) \Phi(L) dL\right) 4\pi R^2 dR,
\end{equation}
where $\Phi(L)$ is the \citet{Hopkins2007} QLF and $L_\mathrm{min}$ is the smallest luminosity quasar that we consider ($L_\mathrm{min}=10^{43}$erg s$^{-1}$ in the B-band, but the overall results do not depend strongly on this choice). The majority of $\Gamma_\mathrm{min}$ comes from cosmologically local sources (within~$\sim75$~cMpc), so neglecting the cosmological redshift of ionizing photons should be a reasonable approximation. The resulting $\Gamma_\mathrm{min}(z)$ will be referred to as the ``minimum" background model in the rest of the paper. We will see in Section~\ref{sect:results} that the minimum model ionization rate is nearly constant over the redshift range we consider.

In our model, the minimum background provides a maximum average opacity for the IGM. To implement the minimum background into our modified cosmological radiative transfer model, we make the following approximation: when determining the opacity $d\tau/dz$ at a given redshift, use the larger of $\GHeII(z)$ and $\Gamma_\mathrm{min}(z)$.\footnote{This discontinuity in the opacity calculation results in a slight redshift discontinuity in the ionizing background evolution, but as mentioned in the text, the mean background we calculate in this regime is unlikely to be physically relevant.} 
The minimum background model is not meant to represent a universe where there is a floor in the ionizing background at any point in space, but rather one where the \emph{average} ionizing background has a floor based on the proximity effects of rare bright sources. This model could be similar to the pre-reionization universe, where the average ionizing background is dominated by expanding ionized bubbles around such sources. While the difference between this highly fluctuating (by construction) background and the pre-reionization universe is subtle, in practice we find that distinction does not matter for our purposes. In the regime where the minimum background model dominates our results, the behaviour of the ionizing background is unlikely to be well-described by globally averaged quantities, so we focus our analysis at redshifts when $\Gamma>\Gamma_\mathrm{min}$. 

The results of this modified cosmological radiative transfer model will be referred to as the "fluctuating" background model in the rest of the paper.

\begin{figure*}
\begin{center}
\resizebox{8cm}{!}{\includegraphics{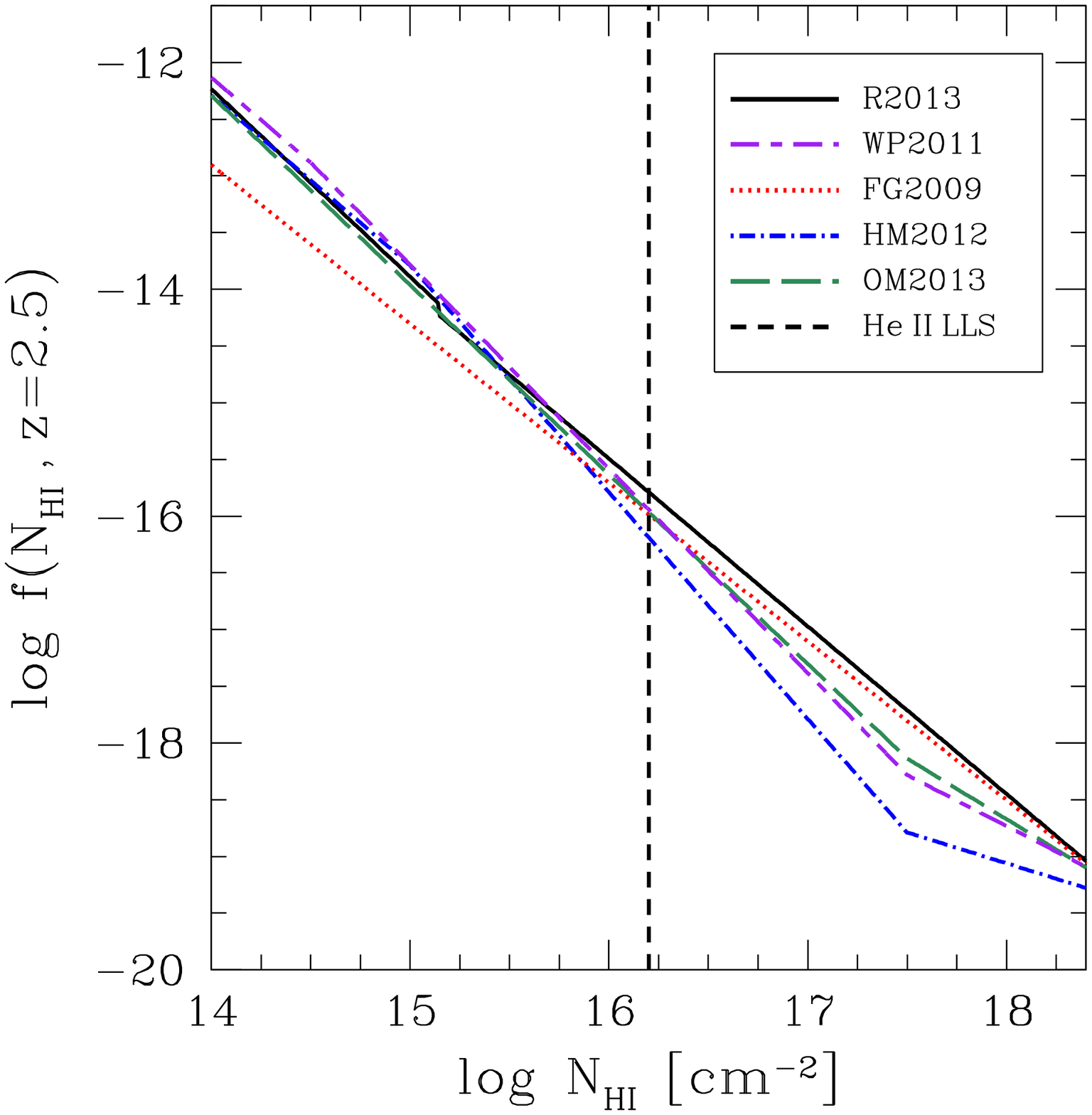}}
\hspace{0.13cm}
\resizebox{8cm}{!}{\includegraphics{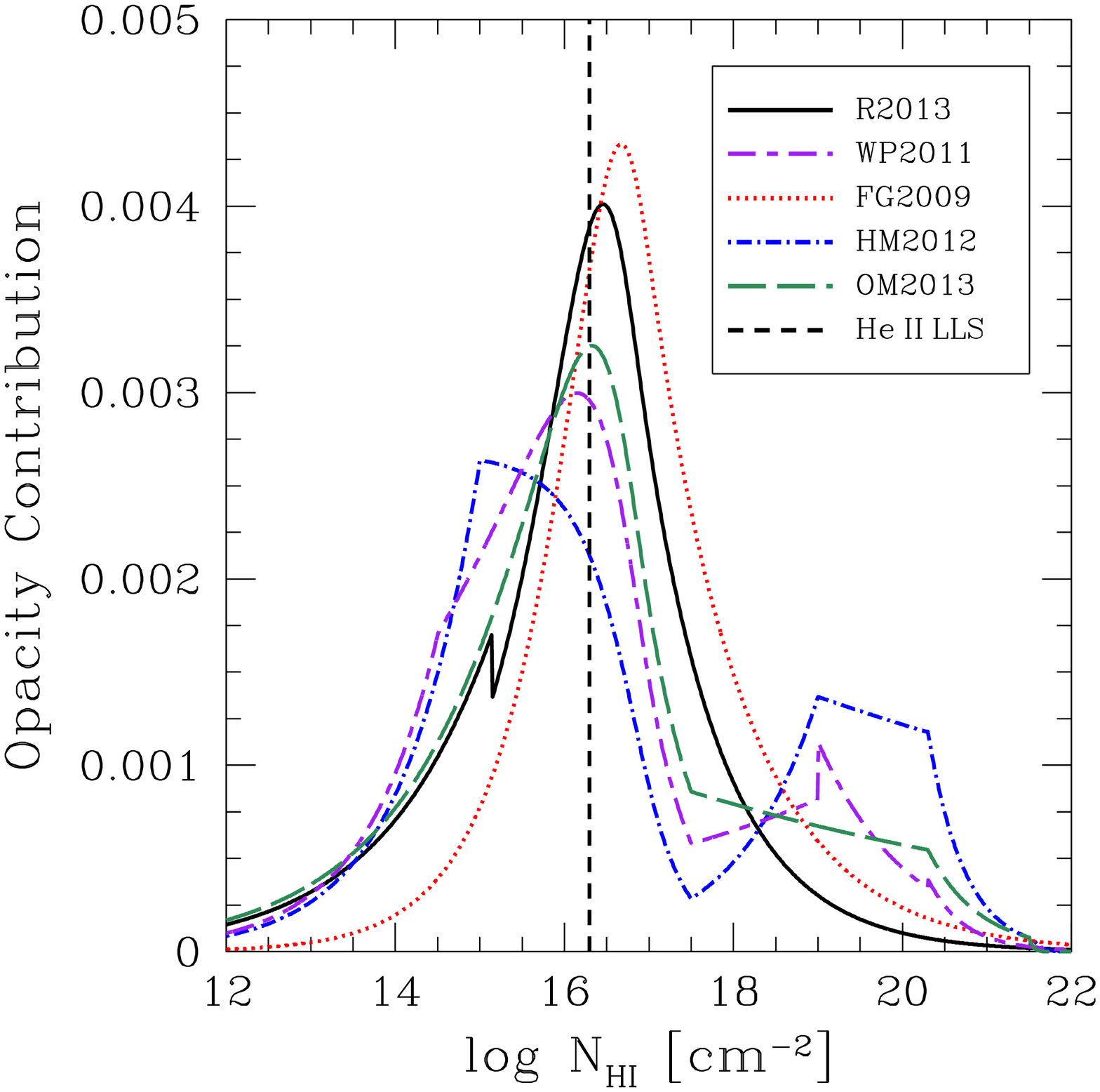}}
\end{center}
\caption{Left: Column density distribution functions $f(N_\HI,z=2.5)$ considered in the text: \citet{Rudie2012b} (solid black), \citet{HM2012} (dot-dashed blue), \citet{FG2009} (dotted red), \citet{OM2012} (long-dashed green), \citet{WP2011} (short-dashed-long-dashed purple), focusing on $N_\HI$ that correspond to the most important He\,\sevensize{\textbf{II}}\small\, absorbers. The vertical dashed line shows the $N_\HI$ corresponding to a He\,\sevensize{\textbf{II}}\small\, ``LLS". Right: Relative contribution to the continuum opacity at $\nu_\HeII$ per log($N_\HI$).}
\label{fig:cddfcontrib}
\end{figure*}

\subsubsection{Summary of Fluctuating Method}

In summary, we calculate the fluctuating background model in the following manner:

\begin{enumerate}
\item Initialize $\GHeII(z)$ and $\lambda_\mathrm{mfp}(z)$ using the standard cosmological radiative transfer approach (equations~\ref{eqn:jnu}--\ref{eqn:Gamma}, \ref{eqn:dtdz}--\ref{eqn:mfp}).
\item Calculate $f(\Gamma)$ as a function of redshift using $\lambda_\mathrm{mfp}(z)$ as input to the method of \citet{Furlanetto2009}.
\item Calculate the average opacity $\langle d\bar{\tau}/dz \rangle$ as a function of redshift using $f(\Gamma)$ (equation~\ref{eqn:fluct}).
\item Calculate $\GHeII(z)$ with equations~\ref{eqn:jnu}--\ref{eqn:Gamma} using $\langle d\bar{\tau}/dz \rangle$ in equation~\ref{eqn:tbar}.
\item Calculate $\lambda_\mathrm{mfp}(z)$ with equations~\ref{eqn:dtdz}--\ref{eqn:mfp}, substituting $\langle d\bar{\tau}/dz \rangle$ for $d\bar{\tau}/dz$ in equation~\ref{eqn:mfp}.
\item Repeat steps (ii)--(v) until $\GHeII(z)$ converges.
\end{enumerate}

\subsection{Model Input Parameters}

Other than our simple model assumptions, the largest sources of uncertainty in our analysis are three  observed parameters: the \HeIIa ionizing emissivity, $\epsilon_\nu$, the \HIa ionization rate, $\Gamma_\HI$, and the neutral hydrogen column density distribution, $f(N_\HI,z)$. In this section, we discuss the range of observed values for these parameters.

\subsubsection{\HeIIa Ionizing Emissivity}\label{sect:emissparam}

We adopt the Lyman limit quasar ionizing emissivity from \citet{HM2012},
\begin{eqnarray}\label{eqn:hmepsilon}
\epsilon_{912}(z) &=& 10^{24.6} \mathrm{erg\,\mathrm{s}^{-1}\,Mpc}^{-3}\,\mathrm{Hz}^{-1} \nonumber \\
&& \times(1+z)^{4.68}\frac{\exp[-0.28z]}{\exp[1.77z]+26.3},
\end{eqnarray}
which is a fit to the integrated $B$-band quasar luminosity function of \citet{Hopkins2007} converted to $\nu_\HI$ by a constant factor,
\begin{equation}\label{h07sed}
L_{\nu_\HI} = L_B \times 10^{18.15} \mathrm{erg\,s}^{-1}\mathrm{\,Hz}^{-1} \left( \frac{L_{\sun}}{L_B}\right) .
\end{equation}
This factor is effectively an estimate of the average quasar spectrum between $\nu_B$ and $\nu_\HI$. For frequencies above the Lyman limit, we assume a power law spectrum with $\epsilon_\nu \propto \nu^{-\alpha}$. For reference, the integrated quasar emissivity given by equation~(\ref{eqn:hmepsilon}) increases by $\sim30\%$ from $z=3$--$2$.

The uncertainty in the \HeIIa ionizing emissivity is a combination of the uncertainty in the \citet{Hopkins2007} quasar luminosity function and the assumed average quasar spectrum. The former is likely to be small, because the integrated quasar $B$-band emissivity at $z\ga2$ comes predominantly from the brightest, and therefore best measured, sources \citep{Hopkins2007}. The latter uncertainty is dominated by the choice of far-UV spectral index $\alpha$. \citet{Telfer2002} find $\alpha = 1.57 \pm 0.17$ for a composite spectrum of 77 radio-quiet quasars, while the composite including an additional 107 radio-loud quasars has $\alpha = 1.76 \pm 0.12$. In contrast, \citet{Scott2004} found that the average spectral index for their sample of 85 sources was considerably harder, $\alpha=0.56^{+0.28}_{-0.38}$. \citet{Shull2012} measured a best-fit spectral index of $\alpha = 1.41 \pm 0.21$ for their sample of 22 sources using \textit{HST/COS}. 

We adopt $\alpha=1.6$ as our fiducial value. Note that, because the \HeIIa Lyman limit $\nu_\HeII = 4\nu_\HI$, a change in the spectral index $\Delta\alpha$ corresponds to a factor of $4^{-\Delta\alpha}$ difference in the emissivity at $\nu_\HeII$.

\subsubsection{\HIa Ionization Rate}
The absorber model in \S~\ref{sect:absmodel} depends on the \HIa ionization rate, $\GHI$. Measurements of $\GHI$ from $z\sim2$--$3$ yield values $\sim0.5-1.0\times10^{-12}$ s$^{-1}$ from flux decrement observations \citep{Rauch1997,Bolton2005,MM2001,FG2008b} or $\sim1.0-3.0\times10^{-12}$ s$^{-1}$ from proximity effect measurements \citep{Scott2000}. The most recent cosmological radiative transfer model by \citet{HM2012} suggests $\GHI \sim 0.8-0.9\times10^{-12}$ s$^{-1}$, but as discussed in the next section, that study may have significantly underestimated the total \HIa opacity of the IGM. We adopt $\GHI = 0.6\times10^{-12}$ s$^{-1}$, a value consistent with the measurements of \citet{FG2008b}, as our fiducial value but consider a range of plausible values.

\subsubsection{Column Density Distribution} \label{sect:CDDF}

The column density distribution of neutral hydrogen $f(N_{\HI},z) = \partial^2N/\partial N_\HI\partial z$ has been measured several times and over a range of redshifts through observations of the \HIa \lya forest. Early observations indicated that the $N_\HI$ distribution is well-fit by a power law of the form $f(N_\HI) \propto N_\HI^{-\beta}$ with $\beta \sim 1.5$ over a wide range of observed column densities ($10^{12} < N_\HI < 10^{22}$ cm$^{-2}$) and redshifts ($z\sim0.2$--$3.5$) \citep{Tytler1987}. Recent studies of \HIa ionizing continuum opacity in stacked quasar spectra at $z\sim2$ and $z\sim4$ suggest a deficit of Lyman limit systems ($10^{17.2} < N_\HI < 10^{19}$ cm$^{-2}$; LLS) and intermediate \HIa column density systems ($10^{15} < N_\HI < 10^{17.2}$ cm$^{-2}$) relative to the canonical single power law model, and several authors have proposed multi-step power law distributions to describe this feature \citep{Prochaska2009,Prochaska2010,WP2011,OM2012,HM2012}. 
\citet{Rudie2012b} performed the largest survey of $10^{12} < N_\HI < 10^{17.2}$ cm$^{-2}$ systems to date for redshifts $z=2.02$--$2.84$ ($\langle z \rangle \sim 2.4$) and found no evidence of the deficit suggested by stacked quasar spectra studies. They found that their measured distribution is well-parameterized by a relatively steep $\beta\sim1.66$ power law for $N_\HI \la 10^{15}$ cm$^{-2}$ and a $\beta\sim1.48$ power law for larger \HIa column densities. The left panel of Figure~\ref{fig:cddfcontrib} shows several of these distributions graphically.

The redshift evolution of the CDDF is usually parameterized by a power law $f(N_\HI,z)\propto(1+z)^\gamma$. However, observationally this $\gamma$ appears to depend on $N_\HI$, implying that the shape of the CDDF evolves with time. The observational constraints on $\gamma$ for $z\ga2$ in the \lya forest regime ($N_\HI < 10^{17.2}$ cm$^{-2}$) are $\gamma\sim2.0$--$3.0$ from line-counting \citep{Kim2002} and measurements of the effective optical depth \citep{FG2008a,DA2008}. The number densities of super-Lyman limit ($10^{19}$ cm $^{-2} < N_\HI < 10^{20.3}$ cm $^{-2}$) and damped \lya ($N_\HI > 10^{20.3}$ cm $^{-2}$) absorbers appear to evolve more slowly with $\gamma\sim1.7$ \citep{OM2007,WP2011} and $\sim1.27$ \citep{Rao2006}, respectively. \citet{Rudie2012b} found that their data were consistent with $\gamma = 2.5$ and $1.0$ for $N_\HI$ below and above $\sim10^{15}$ cm$^{-2}$, respectively.

\begin{figure}
\begin{center}
\resizebox{8cm}{!}{\includegraphics{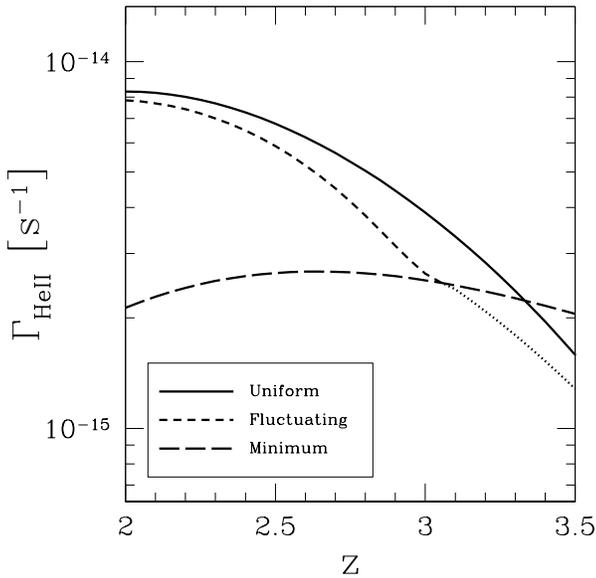}}\\
\end{center}
\caption{Uniform and fluctuating $\GHeII$ in the fiducial model (solid and dashed curves, respectively) and the ``minimum" average ionization rate from isolated quasar profiles (long-dashed). The dotted curve represents the result of the fluctuating model calculation when it is inconsistent (i.e. below) the minimum background model from \S\ref{sect:mingamma}.}
\label{fig:gammafilling}
\end{figure}

\citet{WP2011} and \citet{HM2012} compiled these observations (with the exception of \citealt{Rudie2012b}) and constructed similar multi-step power law CDDFs. The primary difference between the two is the enhanced redshift evolution ($\gamma=3.0$) of \lya forest absorbers in the \citet{HM2012} model compared to the \citet{WP2011} model ($\gamma=2.04$). Both models determine the redshift evolution of the CDDF by comparing to observations of the evolution of the \HIa \lya effective optical depth, which is proportional to $(1+z)^{\gamma+1}$. However, \citet{HM2012} calibrate to the measurements of \citet{FG2008a}, while \citet{WP2011} chose the measurements of \citet{DA2008}. It is unclear why such a difference exists in the effective optical depth evolution measured by these two groups, but it does not significantly affect our results.

In the following sections, we use the broken power-law CDDF from \citet{Rudie2012b} as our fiducial model. Their model represents the first solid measurement of intermediate \HIa column density absorbers that are critical to the \HeIIa ionizing opacity, and is consistent with measurements of the \HIa \lya effective optical depth (G. Rudie, priv. comm.). However, as the following sections will show, our choice of CDDF does not have significant implications for our final results, given the overall uncertainty in the amplitude of the ionizing background.

\section{Evolution of the Ionizing Background} \label{sect:results}

\subsection{The Ionizing Background With Uniform Emissivity}

The solid curve in Figure~\ref{fig:gammafilling} shows how the \HeIIa ionization rate ($\GHeII$) evolves in our \emph{uniform} fiducial model, ignoring fluctuations in the ionizing background. The uniform background model results in a steeply evolving ionizing background from $z\sim3$--$2$, with an ionization rate that increases by a factor of $\sim2$ over that range before flattening out substantially at later times. In the following sections, we discuss how variations in the input parameters affect this result.

\begin{figure}
\begin{center}
\resizebox{8cm}{!}{\includegraphics{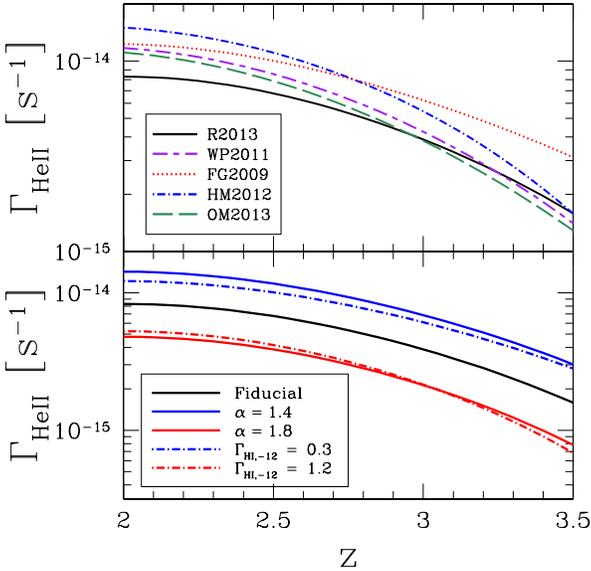}}\\
\end{center}
\caption{Top: $\GHeII$ in the uniform background model calculated for CDDFs from Figure \ref{fig:cddfcontrib}. Bottom: Effect of assumed average quasar spectrum shortward of 912 \AA\,, given by $F_\nu\propto\nu^{-\alpha}$ (solid curves), and assumed (constant) $\Gamma_\HI$ (dot-dashed curves).}
\label{fig:alphaghicddf}
\end{figure}

\subsubsection{Column Density Distribution} \label{sect:cddfresults}

We considered a variety of CDDFs in our model. The top panel of Figure~\ref{fig:alphaghicddf} shows the uniform ionizing background calculated with CDDFs used in earlier ionizing background calculations by \citet{HM2012} and \citet{FG2009}, the direct measurement at $\langle z \rangle\sim2.4$ by \citet{Rudie2012b}, and indirect extrapolation from higher redshift opacity measurements \citep{WP2011,OM2012}. In general, despite the significant differences between CDDFs apparent in the left panel of Figure~\ref{fig:cddfcontrib}, the evolution of the uniform background from $z=3$--$2$ is fairly insensitive to the CDDF. The most significant differences are due to the different redshift evolution of the CDDFs, which is not very well constrained.

The right panel of Figure~\ref{fig:cddfcontrib} shows the relative contribution to the ionizing continuum opacity at the \HeIIa edge as a function of $N_\HI$. Most of the opacity is due to \HeIIa ``LLSs" with $N_\HeII \sim \sigma_\HeII^{-1}$, but the multi-step power law CDDFs have an increased contribution from relatively low $N_\HI$ ($\la 10^{15}$ cm$^{-2}$) absorbers compared to the shallow power law \citet{FG2009} CDDF. The \HIa column density corresponding to the peak \HeIIa opacity contribution varies from $10^{15}$ to $10^{16.7}$ cm$^{-2}$ depending on the shape of the CDDF.

Figure~\ref{fig:alphaghicddf} also shows that the normalization of $\GHeII$ depends sensitively on the total opacity calculated from the CDDF, which can vary significantly between models. If $\GHeII$ were accurately measured near $z\sim2$, that measurement could in principle be used to help distinguish between models. However, measuring $\GHeII$ directly is extremely difficult, and as shown in the following sections, the other model parameters can be adjusted to produce similar differences in the normalization. For example, measurements of both the $\etath$ parameter and $\GHI$ could potentially be used to constrain acceptable normalizations of $\GHeII$ (because the expected \HeIIa Ly$\alpha$ opacity in the IGM depends strongly on the value of the former parameter; see equation~\ref{eqn:etath}), but the current constraints on these parameters are too weak, and the degeneracies are too strong, to distinguish between the models presented in this and the following sections. 

\subsubsection{Quasar Spectrum} \label{sect:emissresults}

To assess the effect of choosing different average far-ultraviolet quasar spectral indices, we fix the \HIa Lyman limit emissivity given by equation~\ref{eqn:hmepsilon} and scale to \HeIIa ionizing photons by $\epsilon_\nu \propto \nu^{-\alpha}$. The solid curves in the bottom panel of Figure~\ref{fig:alphaghicddf} show how the range of observed values of the quasar spectral index $\alpha$ affects the \HeIIa ionization rate. A harder spectrum, which produces more ionizing photons at $\nu_\HeII$, results in a higher ionization rate. Fixing the emissivity at $\nu_\HeII$ and changing the spectral index has very little effect on the resulting $\GHeII$. In contrast, we find that $\GHeII$ changes more strongly than linearly with $\epsilon_\HeII$; this is because the absorber structure changes with the ionizing background (and hence the emissivity). In general, as $\GHeII$ increases, the \HIa column density corresponding to a \HeIIa LLS increases. Since $N_\HI f(N_\HI,z)$ is a decreasing function of $N_\HI$, the number density of \HeIIa LLSs, and thus the overall opacity, decreases. This behaviour is similar to the emissivity-$\Gamma$ feedback studied by \citet{McQuinn2011}. The redshift evolution of the background is affected by $\alpha$ as well, but the effect is subtle.

\subsubsection{\HIa Ionization Rate} \label{sect:ghiresults}

The dot-dashed curves in the bottom panel of Figure~\ref{fig:alphaghicddf} show how the \HeIIa ionization rate is affected by the assumed value of $\GHI$. The effect is similar to changing the number of \HeIIa ionizing photons,  because both parameters modulate the ratio of \HeIIa to \HIa in absorbers. While the decrease in \HeIIa opacity with an increasing number of \HeIIa ionizing photons is straightforward in principle, the relationship between $\GHI$ and $\GHeII$ is more subtle. Consider an optically thin absorber: if $\GHI$ decreases, the amount of \HIa in a fixed physical structure will increase while the amount of \HeIIa stays the same. This shift of the \HIa column density corresponding to a \HeIIa LLS causes $\GHeII$ to change with $\GHI$: if $\GHI$ is larger, the $N_\HI$ corresponding to a \HeIIa LLS will decrease, so \HeIIa LLSs will be more numerous and the overall \HeIIa opacity will increase. $\GHI$ appears to affect the redshift evolution more strongly than $\alpha$.

\begin{figure}
\begin{center}
\resizebox{8cm}{!}{\includegraphics{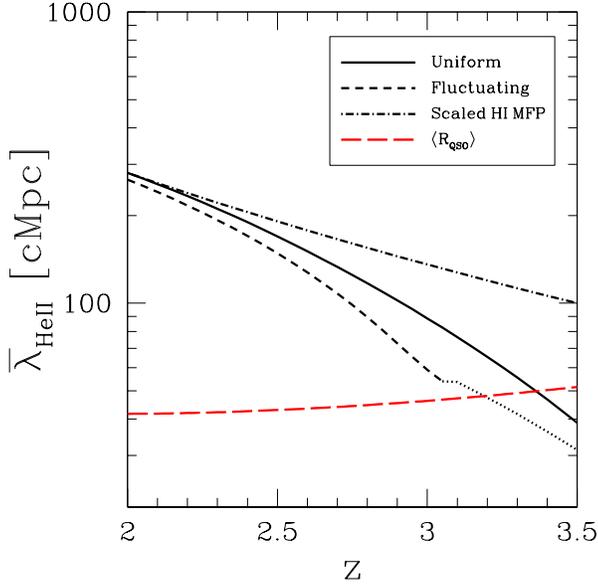}}\\
\end{center}
\caption{Evolution of the He\,\sevensize{\textbf{II}}\small\, ionizing photon mean free path with redshift (black), evaluated at the average ionizing energy, for the uniform (solid) and fluctuating (dashed) models. The dotted curve represents the results of the fluctuating model when it is inconsistent with the minimum expected background from \S\ref{sect:mingamma} as in Figure~\ref{fig:gammafilling}. The evolution of the hydrogen ionizing photon mean free path is shown as the dot-dashed curve, scaled to the He\,\sevensize{\textbf{II}}\small\, mean free path at $z = 2$. The red long-dashed curve shows the average separation between luminous ($\nu_BL_B > 10^{11} L_{\sun}$) quasars given by the \citet{Hopkins2007} QLF.}
\label{fig:mfpevol}
\end{figure}

\subsubsection{Mean Free Path} \label{sect:mfpresults}

The solid curve in Figure~\ref{fig:mfpevol} shows the evolution of $\bar{\lambda}_\HeII$ in the uniform model.  We also show how $\lambda_\HI$ increases with cosmic time (dotted curve); for ease of comparison we scale this curve to $\bar{\lambda}_\HeII$ at $z = 2$. In contrast to the power-law evolution of $\lambda_\HI$ (described by equation~\ref{eqn:lambda}), $\bar{\lambda}_\HeII$ evolves much faster than a simple power law.  

The evolution of the mean free path at the \HeIIa ionizing edge in our fiducial model is well-approximated by a power law with an index that itself evolves as a power law,
\begin{eqnarray}\label{eqn:lambdafit}
\lambda_\HeII &\sim& 188\,\,\mathrm{comoving\,Mpc}\,\,\times \left(\frac{1+z}{3}\right)^{\zeta(z)} \\
\zeta(z) &=& -2.41 \times \left(\frac{1+z}{3}\right)^{1.92}.
\end{eqnarray}
This fit differs by no more than $\sim3\%$ from our full numerical calculations over the redshift range $z=2$--$3.8$, but we caution the reader that the systematic uncertainties from our model input parameters are much, much larger than this.  We also caution the reader against using this fit at $z \ga 3.4$, where fluctuations in the ionizing background \emph{must} be included (see below).

Because $\lambda_\HeII$ is linked to $\GHeII$ through the absorber structure prescription, it evolves more quickly than $\lambda_\HI$. That is, increasing the mean free path increases the ionizing background, which will then increase the \HIa column density at which \HeIIa becomes optically thick, which in turn increases the mean free path, etc. This feedback effect is the fundamental source of the rapid evolution we see in $\GHeII$. (In fact, one could argue that it is strange that such rapid evolution does \emph{not} occur in $\GHI$; see \citealt{McQuinn2011}.)

The dependence of the mean free path on frequency is a function of the logarithmic slope of the CDDF, $\lambda_\mathrm{mfp} \propto \nu^{3(\beta-1)}$ (equation~\ref{eqn:lambda}). The \HeIIa CDDF is not precisely defined in our model, but a mapping of our fiducial \HIa CDDF through our absorber prescription results in $\beta_\HeII\sim1.43$ for the absorbers that contribute the bulk of the opacity near the \HeIIa ionizing edge ($10^{14.5} \la N_\HI \la 10^{17.0}$ cm$^{-2}$ as in Figure~\ref{fig:cddfcontrib}), and consequently $\lambda_\HeII \propto \nu^{1.3}$ for $1 \leq \nu/\nu_\HeII \la 2.5$. $\bar{\nu}/\nu_\HeII\sim1.37$ is typical for our fiducial model, so $\bar{\lambda}_\HeII/\lambda_\HeII\sim1.51$.

\subsubsection{Recombination Photons} \label{sect:recresults}

The fractional contribution of recombination emission to $\GHeII$ is fairly minor. In the absence of quasars, but with the opacity as a function of redshift fixed to the uniform model, recombination photons alone produce an ionization rate about $\sim7$--$15\%$ of the fiducial value. However, because the absorber population is sensitive to the emissivity (as in \S~\ref{sect:emissresults}), the relative difference between $\GHeII$ calculated with recombination emission and $\GHeII$ calculated without recombination emission is larger ($\sim20$--$40\%$). While \citet{FG2009} found that including recombination emission increased $\GHeII$ by only $\sim10\%$, Figure~\ref{fig:cddfcontrib} shows that their CDDF has a significant deficit of the optically thin ($N_\HI \la 10^{16}$ cm$^{-2}$) systems that contribute most of the recombination emissivity. 

In simple models of the reionization process, it is conventional to describe the enhanced recombination rate of ionized species $n_i$ due to an inhomogeneous IGM through the so-called clumping factor, $C=\langle n_i n_e \rangle / (\langle n_i \rangle \langle n_e \rangle)$. Usually, this is estimated from simple phenomenological arguments or from the density structure in numerical simulations.  However, these approaches are not entirely satisfactory, as the clumping factor should incorporate information that depends on the distribution of ionized and neutral patches.  For example, recombinations that occur inside of dense, self-shielded systems do not produce photons that can ionize the IGM, as the resulting photons are trapped within the systems.

With our detailed model, we can estimate this factor for \HeIIIa self-consistently (given a model for the emitting and absorbing populations) by explicitly following the fraction of recombinations that occur inside of self-shielded systems. In particular, we have
\begin{equation}
C_\mathrm{eff} = \frac{\int_{\nu_\HeII}^\infty \epsilon_{\nu,\mathrm{rec}}/(h\nu) d\nu}{(\alpha_\HeII^A-\alpha_\HeII^B)\langle n_\HeIII \rangle \langle n_e \rangle },
\end{equation}
which describes the effective recombination rate after correcting for self-absorption of ionizing recombination photons within the emitting clouds relative to a uniform IGM. In our fiducial uniform model, $C_\mathrm{eff}$ increases from $C_\mathrm{eff}\sim1$ at $z=3.5$ to $C_\mathrm{eff}\sim4$ at $z=2$.

\subsection{The Ionizing Background Including Fluctuations}\label{sect:fluctresults}

It is instructive to compare the mean free path from the preceding section to the average separation between the primary sources of ionizing photons, bright quasars with $\nu_B L_B > 10^{11} L_{\sun}$.  We calculate the number density of the bright quasars by integrating the \citet{Hopkins2007} luminosity function over this luminosity range and estimating their average separation by $\langle R \rangle \sim n^{-1/3}$. The long-dashed red curve in Figure~\ref{fig:mfpevol} shows this separation; $\langle R \rangle \sim 45$ Mpc is a good approximation for the entire redshift interval from $z\sim2$--$3$.

\begin{figure}
\begin{center}
\resizebox{8cm}{!}{\includegraphics{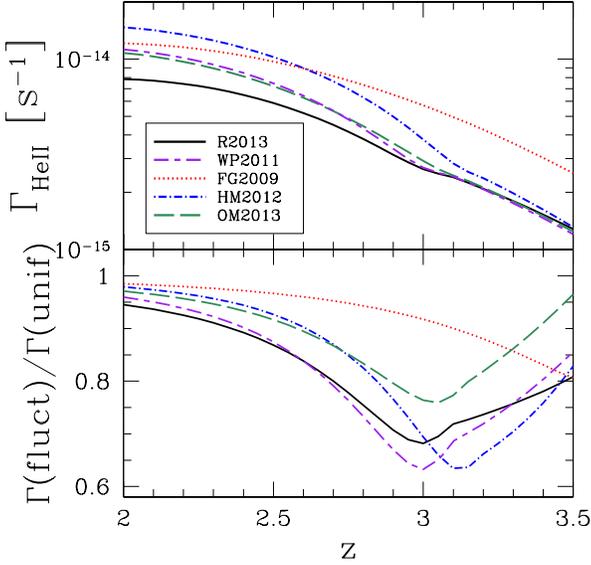}}\\
\end{center}
\caption{Top: Fluctuating model ionization rate for the same CDDFs as Figure~\ref{fig:cddfcontrib}. Bottom: Ratio of the fluctuating and uniform model ionization rates.}
\label{fig:cddfratios}
\end{figure}

When the mean free path is similar to the average source separation, fluctuations in the background contribute a substantial opacity excess. The dashed curve in Figure~\ref{fig:gammafilling} shows the effect of these fluctuations on the ionizing background.  Figure~\ref{fig:cddfratios} shows that, compared to the uniform model, the fluctuating background model exhibits a $\sim20$--$40\%$ dip at $z \sim3$--$3.2$ for our fiducial input parameters and various CDDFs from \S~\ref{sect:cddfresults}. The evolution of all the CDDF models, with the exception of the shallow slope model from \citet{FG2009}, is very similar. The ``feedback" effect between the opacity and the ionizing background is weaker for shallower CDDF slopes (e.g. \citealt{McQuinn2011}), so the net effect of fluctuations is smaller in the \citet{FG2009} model. The ``recovery" of the fluctuating model at higher redshift relative to the uniform model is due to our minimum ionization rate approximation from Section~\ref{sect:mingamma} which limits the effective opacity to ionizing photons.

We note that there are two related sources for the differences between the curves in Figure~\ref{fig:cddfratios}: the shape of the column density distributions and the normalization of the ionizing background. Leaving the CDDF shape fixed, different choices for the ionizing emissivity result in very similar shapes to those in Figure~\ref{fig:cddfratios}, though the redshift above which the minimum background is larger than the fluctuating background will shift depending on the relative normalization of the two models. The more subtle differences in the shapes of the curves are due to variations in the shape of the CDDF; this is most dramatically seen by the dotted curve (which is only below the minimum background model at $z \ga 3.8$).

The most important effect of including fluctuations is to induce a more rapid increase in $\GHeII$ with cosmic time.  Consider a region with a smaller than average emissivity. In that region, the ionizing background will also be smaller, so each absorber will be more optically thick and the mean free path will be smaller.  This will further decrease the ionizing background, etc. In a realistic model of the distribution of $\GHeII$ in the presence of quasars, most of the volume of the universe has an ionizing background a few times \emph{below} the universal average (to compensate for the very brightly illuminated, but small, regions around quasars; see Figure~\ref{fig:fgamma}). Thus, the average opacity through the universe is higher, decreasing the resulting mean ionizing background.

The turndown from the uniform model is thus a straightforward and robust prediction of our fluctuating background model, though its magnitude depends on the CDDF. At higher redshift ($z \ga 3$), it is clear that the \HeIIa ionizing background evolution should no longer be described by a cosmological radiative transfer model without properly taking into account the effect of localized transparent regions around sources. Our simple analytic model for the minimum ionization rate from isolated quasars, the minimum background model, should represent a fairly strict lower limit to the ionizing background in the post-reionization (i.e. ionization equilibrium) limit. If this is indeed the case, one might expect the volume-averaged ionization rate to evolve \emph{more slowly} at higher redshift ($z \ga 3.1$) than predicted by standard cosmological radiative transfer. While our minimum model neglects a diffuse partially-neutral component to the IGM that should exist prior to the completion of \HeIIa reionization, this slower evolution is consistent with the \HeIIa reionization simulations of \citet{McQuinn2009} and with expectations from hydrogen reionization \citep{Furlanetto2009b}. In both cases, the ionizing background is reduced to a set of independent ``proximity zones" (though for different reasons), with the mean background depending principally on the filling factor of these regions. 

These calculations show that the ionizing background can evolve very rapidly at $z \la 3$, \emph{even without any assumptions about an evolving \HeIIa fraction}. The precise degree of evolution is uncertain, but it is at least a factor of a few--even in the standard uniform emissivity model--and likely nearly a factor of five when fluctuations are included. In other words, even without late \HeIIa reionization, we should see a rapid increase in the intensity of the metagalactic radiation field.  This evolution is in stark contrast to observations of the \HIa ionization rate, which appears to be roughly constant from $z\sim2$--$4$; this difference is most likely due to the increasing influence (towards higher redshift) of star-forming galaxies (as opposed to quasars) to the \HIa ionizing emissivity. We will consider the observable implications of this conclusion in the following section.

For $z \la 3$, the mean free path at the \HeIIa ionizing edge in the fluctuating background model is well-characterized by a similar power law within a power law as the uniform model (equation~\ref{eqn:lambdafit}),
\begin{eqnarray}\label{eqn:lambdafitfluct}
\lambda_\HeII &\sim& 178\,\,\mathrm{comoving\,Mpc}\,\,\times \left(\frac{1+z}{3}\right)^{\zeta(z)} \\
\zeta(z) &=& -2.64 \times \left(\frac{1+z}{3}\right)^{2.61}.
\end{eqnarray}
The primary difference between the uniform and fluctuating background fits is the larger power law index of $\zeta(z)$, a consequence of faster ionizing background evolution. As discussed previously, the mean free path of average energy ionizing photons that we use in the fluctuating background calculation is somewhat larger:
\begin{eqnarray}
\bar{\lambda}_\HeII &\sim& 266\,\,\mathrm{comoving\,Mpc}\,\,\times \left(\frac{1+z}{3}\right)^{\zeta(z)} \\
\zeta(z) &=& -2.62 \times \left(\frac{1+z}{3}\right)^{2.38}.
\end{eqnarray}

\section{Effective Optical Depth} \label{sec:taueff}

To gauge the observable import of our results, we will briefly consider how they manifest in the evolution of the IGM opacity to far-ultraviolet photons. \HeIIa \lya absorption has been measured in far-ultraviolet spectra from $z \sim 2$--$4$ (\citealt{DF2009} and references therein; \citealt{Worseck2011,Syphers2011,Syphers2012,SS2012}). We will compare to the most basic observable from the resulting forest of observed absorption features, the average optical depth $\tau_\mathrm{eff}$ for the \HeIIa \lya transition. We use two different methods to predict $\tau_\mathrm{eff}$: a semi-analytic model using a gas density probability distribution $P(\Delta)$ as in \citet{DF2009}, and a direct integration of the \HeIIa \lya opacity from the \HIa CDDF and our absorber structure prescription.

\begin{figure}
\begin{center}
\resizebox{8cm}{!}{\includegraphics{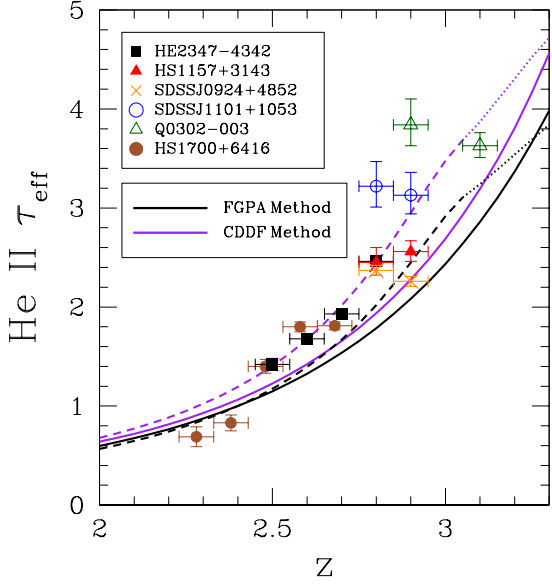}}\\
\end{center}
\caption{Effective optical depth for the uniform (solid), fluctuating (dashed), ``plateau" (dotted), and ``minimum" (dash-dotted) models, using the FGPA (black) and CDDF (purple) methods. The points are $\Delta z=0.1$-binned effective optical depth data for various quasar sightlines from \citet{SS2012} ( (HS1700+6416, filled brown circles; excluding metal absorption) and \citet{Worseck2011} (G. Worseck, priv. comm.; HE2347-4342, black squares; HS1157+3143, filled red triangles; SDSSJ0924+4852, orange crosses; SDSSJ1101+1053, open blue circles; Q0302-003, open green triangles).}
\label{fig:opdepth}
\end{figure}

Under the assumptions of a highly-ionized universe in ionization equilibrium, line opacity dominated by zero-width optically thin absorbers, and a power-law temperature-density relation $T = T_0\Delta^{1-\gamma_{d}}$, the \HeIIa Gunn-Peterson optical depth can be expressed as \citep{DF2009}
\begin{eqnarray}\label{eqn:taugp}
\tau_\mathrm{GP} &\simeq& 13.6\kappa \left( \frac{\Gamma_\HeII}{10^{-14}\mathrm{\,s}^{-1}} \right) ^{-1} \left( \frac{T_0}{10^4\mathrm{\,K}} \right) ^{-0.7} \left( \frac{\Omega_bh^2}{0.0241} \right) ^2 \nonumber \\
&& \times \left( \frac{\Omega_mh^2}{0.142} \right) ^{-1/2} \left( \frac{1+z}{4} \right) ^{9/2}\Delta^{2-0.7(\gamma_{d}-1)} .
\end{eqnarray}
This ``fluctuating Gunn-Peterson approximation" (FGPA; \citealt{Weinberg1997}) relates the continuum optical depth to the local overdensity $\Delta$ and ionization rate $\GHeII$. The systematic uncertainty in $\tau_\mathrm{GP}$ due to the above simplifications is collapsed into a normalization constant $\kappa$, which we calibrate to (one of) the observations. We assume an isothermal temperature-density relation ($\gamma_d = 1$) for simplicity, but this does not affect our results significantly. Assuming the gas density probability distribution given by \citet{MHR2000}, we then calculate $\tau_\mathrm{eff}$ by integrating over the density and ionization rate distributions:
\begin{equation}\label{eqn:taueff}
e^{-\tau_\mathrm{eff}} = \int_0^\infty d\Gamma f(\Gamma) \int_0^\infty d\Delta e^{-\tau_\mathrm{GP}(\Gamma,\Delta)} P(\Delta).
\end{equation}
We normalized the FGPA results for the uniform and fluctuating models to produce an optical depth of $\tau=1.0$ at $z=2.4$ to roughly match observations \citep{Worseck2011,SS2012} when the expected variation between sightlines is small. These normalizations require $\kappa=1.56$ and $\kappa=1.28$ (equation~\ref{eqn:taugp}) for the uniform and fluctuating models, respectively.

An alternative method to calculate $\tau_\mathrm{eff}$ is to directly integrate the \HeIIa \lya opacity from the CDDF. The only additional information needed is the distribution of line widths, provided by the Doppler parameter $b$.  In this method, $\tau_\mathrm{eff}$ is given by \citep{Zuo1993}
\begin{eqnarray}
\tau_\mathrm{eff} &=& \frac{1+z}{\lambda_{\HeII,\mathrm{Ly}\alpha}} \nonumber \\
&& \times \int_{N_{\HI,\mathrm{min}}}^{N_{\HI,\mathrm{max}}} dN_\HI \int_0^\infty db f(N_\HI,b) W(N_\HI,b) ,
\end{eqnarray} 
where $W(N_\HI,b)$ is the \HeIIa \lya equivalent width of an absorber with Doppler parameter $b$ and $f(N_\HI,b)$ is the joint distribution of $N_\HI$ and $b$. We assume that $N_\HI$ and $b$ are uncorrelated and that the distribution of $b$ is a Dirac-delta function at $b = 30$ km s$^{-1}$, a representative approximation for \HIa \lya forest systems \citep{Kim2001}. In this method we do not subject the resulting optical depth to any extra normalization.

The results of the FGPA and CDDF methods are shown in Figure~\ref{fig:opdepth}. Both methods demonstrate that steep evolution of $\GHeII$ naturally leads to steep evolution in the observed $\tau_\mathrm{eff}$. The addition of fluctuations further accelerates the evolution. The results for different input parameters ($\alpha$, $\GHI$, CDDF) are largely the same in the FGPA method when normalized at $z=2.4$. In contrast, the CDDF method depends sensitively on $\etath\propto\GHI/\GHeII$ (equation~\ref{eqn:etath}), which can differ by a factor of a few between models. Thus, for a given CDDF, the \HeIIa optical depth places a joint constraint on $\GHI$ and $\alpha$, subject to the uncertainties inherent in our cosmological radiative transfer model.

For context, we also show measured $\tau_{\rm eff}$ points in Figure~\ref{fig:opdepth} from \citet{SS2012} and \citet{Worseck2011}. These two works determine the effective optical depth by measuring the transmission uniformly across the redshift interval ($\tau_\mathrm{eff}=-\mathrm{ln}\,\langle F \rangle$) instead of averaging transmission from sparse redshift coverage provided by past works \citep{DF2009} or averaging pixel optical depths \citep{Shull2010}. It is interesting that our models -- which explicitly ignore \HeIIa reionization -- match the evolution in the observed optical depth rather well.  Additionally, the fluctuating background models appear to match the observations more closely than the uniform models, especially at $z\ga2.7$ where the observed optical depth evolution is very steep. Our result demonstrates that the observed trend in and of itself does not \emph{require} the \HeIIa fraction to evolve, although it also does not rule out such evolution.

Unfortunately, our models do not explicitly describe how the integrated $\tau_\mathrm{eff}$ should vary at the same redshift along different lines of sight, even when averaged over large path lengths. This is because our model assumes that the high and low $\Gamma$ regions are distributed perfectly randomly, without the spatial correlations between them that are essential to understanding the observed averages \citep{FD2010}. Hydrodynamic simulations by \citet{McQuinn2009} and semi-analytic models by \citet{FD2010} have described spatial variations in $\tau_\mathrm{eff}$. Interestingly, the well-studied spectrum of HE 2347--4342 \citep{Reimers1997,Kriss2001,Zheng2004,Shull2004,Shull2010} shows regions of high optical depth that appear to require large swathes of \HeIIa at $2.7\la z \la 2.9$. We therefore emphasize that our models do not demand that \HeIIa reionization be over by $z \sim 3$; they instead demonstrate that, with respect to the evolution of the mean opacity, it is not required.

\section{Discussion}
Our model for background fluctuations increases the average opacity of the IGM when the mean free path is comparable to the separation between bright sources. This effect is primarily due to the skewness of $f(\Gamma)$ towards lower $\Gamma$ as the mean free path decreases (as in Figure~\ref{fig:fgamma}). While the effect of our fluctuations prescription on the ionizing background is relatively small, it predicts a steep increase in the ionizing background when the background transitions from being dominated by local sources to a smoother background with contributions from distant sources.

\begin{figure}
\begin{center}
\resizebox{8cm}{!}{\includegraphics{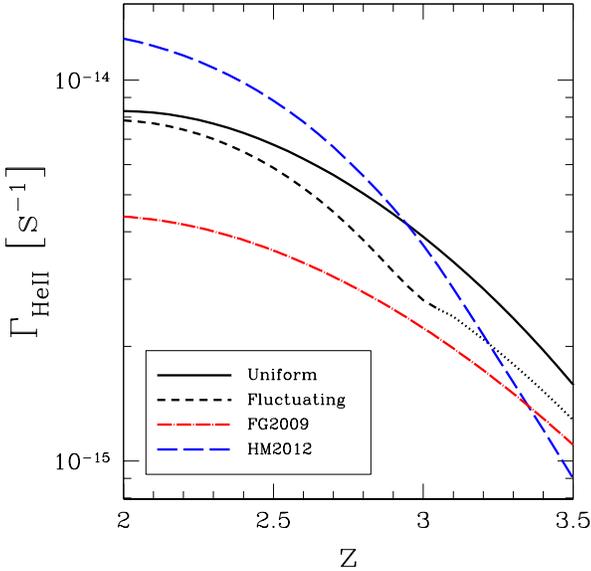}}\\
\end{center}
\caption{Uniform (solid black) and fluctuating (dashed black) He\,{\sevensize{\textbf{II}}} ionization rate from this work compared to the models from \citet{HM2012} (long-dashed blue) and \citet{FG2009} (long-dash-dotted red).}
\label{fig:pastmodels}
\end{figure}

\subsection{Comparison to past theoretical work}

Figure \ref{fig:pastmodels} shows how our model compares to a pair of recent ionizing background calculations by \citet{HM2012} and \citet{FG2009}.

\citet{FG2009} used a single power-law CDDF with $\beta=1.4$ and $\gamma=1.5$ that severely underestimates the number of low-density Ly$\alpha$ forest absorbers compared to recent observations (see the left panel of Figure \ref{fig:cddfcontrib}) and evolves more slowly than implied by Ly$\alpha$ forest measurements \citep{Kim2002}. Because their CDDF severely underestimates the \HIa opacity of the IGM from sub-LLS absorbers, they were forced to renormalize the quasar emissivity of ionizing photons at the hydrogen ionizing edge by a factor of $0.36$ to match their measured $\GHI\sim0.5\times10^{-12}$ s$^{-1}$ \citep{FG2008b}, and thus their $\GHeII$ is normalized somewhat lower as well. Their $\GHeII$ evolves at a similar rate to our fiducial uniform background model.

\citet{HM2012} used a CDDF that evolves more rapidly with redshift than our fiducial model and calculated a $\GHI$ that peaks at $z\sim2$ and declines slowly towards higher redshift. They also used a different fitting form for the structure of IGM absorbers. In their fit they more accurately approximated the average ionization rate within absorbers, which resulted in a more accurate fit to $\eta$ at large \HIa column densities. However, as mentioned previously in \S~\ref{sect:absmodel}, those high $N_\HI$ systems do not contribute a substantial fraction of the opacity near the \HeIIa edge, and thus our approximation should not significantly affect our results.

\subsection{Fluctuating Model Caveats}

Other than the general simplifications necessary to invoke the cosmological radiative transfer model, our parameterization of the fluctuations in the background is an ad hoc addition to a model designed for a medium with a uniform emissivity. In this section, we describe the primary uncertainties with such an approach.

First of all, we may not have accurately captured the extent and character of the fluctuations. Spatial correlations in the ionizing background exist due to the large proximity regions of the primary sources (as seen in the minimum model of \S~\ref{sect:mingamma}). It is possible that a full characterization of the ionizing background fluctuations including proximity effects would negate the need to separately consider the minimum background due to isolated sources, though obviously such an effort is different. Additionally, the massive hosts of these luminous quasars are clustered, which will increase the amplitude of the fluctuations. However, the proximity zones of the quasars are so large, and the quasars so rare, that stochastic variations dominate over large-scale clustering in all reasonable scenarios anyway \citep{Dixon2012}. The absorbers also show some clustering \citep{Rudie2012,Rudie2012b} which will modulate the metagalactic radiation field (although likely only modestly).

Other obvious sources of additional fluctuations in the ionizing background -- over and above those from the discrete sources -- include radiative transfer effects (e.g. ``shadows" behind optically thick regions as in \citealt{TM2007}) and collisional ionization in superheated shocks \citep{Muzahid2011}. Of course, incomplete \HeIIa reionization may leave opaque patches of \HeIIa that would introduce severe fluctuations \citep{McQuinn2009, FD2010} which have possibly been observed recently \citep{Zheng2004,Shull2010,Worseck2011}. We have explicitly ignored this possibility here so as to consider the evolution of the ionizing background in the absence of such effects.

We also treat recombinations only approximately. We include recombination emission in our fluctuating model calculation in the same way as in the uniform model, by simply adding to the pre-existing quasars' emissivity. It therefore implicitly has the same source distribution, while in fact it will be more uniform than the point-like quasars because it is distributed throughout the IGM.  On the other hand, recombination emission in low $\GHeII$ regions will be weaker, and much of the emission from high $\GHeII$ regions (i.e. near bright quasars) will not travel much beyond those quasar proximity regions before redshifting below $\nu_\HeII$ ($\la 30$ Mpc; \S~\ref{sect:recombs}), so its effect on $f(\Gamma)$ should be fairly minor.

\section{Conclusion}

We have calculated the \HeIIa ionizing background using a cosmological radiative transfer model that takes into account the latest constraints on quasar and IGM source properties. In our uniform background model, which closely mimics previous work \citep{Fardal1998,FG2009,HM2012}, we found that the \HeIIa ionization rate, $\GHeII$, and the mean free path of \HeIIa ionizing photons should both evolve significantly during the time after \HeIIa reionization ($z\sim2-3$). However, at $z\sim3$, the mean free path of \HeIIa ionizing photons is comparable to the average distance between the bright quasars that contribute most of the ionizing emissivity. While previous work investigated how this effect introduces fluctuations in the ionizing background \citep{Fardal1998,MW2003,Furlanetto2009}, its implications for the \emph{mean} ionizing background itself have not been studied in detail until now.

We investigated for the first time how these fluctuations can affect the evolution of the mean background. We incorporated the distribution $f(\Gamma)$ into our cosmological radiative transfer model by averaging the opacity to \HeIIa ionizing photons over it.  However, that procedure still models the emission as diffuse sources rather than point-like quasars, so we supplemented it with a physical model that accounts for the decreased average opacity at high redshift by considering \emph{isolated} transparent zones around individual quasars. Including that model, our results showed that the fluctuating background introduces another source of opacity which causes the ionization rate to decrease by a factor of $\sim30\%$ at $z\sim3.1$ relative to the uniform background calculation. For $z\ga3.1$, the cosmological radiative transfer model predicts a mean background below the minimum model, suggesting that it is no longer adequate to properly model the evolution of the \HeIIa ionizing background at those redshifts.

As an example of the utility of our ionizing background model, we used the resulting ionization rate to estimate the evolution of the \HeIIa \lya effective optical depth, $\tau_\mathrm{eff}$. Rapid evolution at $z\ga2.5$, similar to that seen in observations, appears to be a natural consequence of a steeply evolving ionization rate. The addition of fluctuations improves our model's resemblance to the observed $\tau_\mathrm{eff}$ evolution somewhat, though systematic uncertainties in the data analysis make a detailed comparison difficult.

We note that our model does not incorporate \HeIIa reionization: that is, we assume that the \HeIIa fraction is very small throughout the IGM.  We have therefore shown that reionization is not the \emph{only} possible cause of a rapidly evolving ionizing background.  Instead, the interaction between the (slowly) increasing emissivity and the (slowly) evolving IGM clumpiness can feed back on each other, strongly amplifying the evolution of the ionizing background. Such evolution is naively predicted by simple models \citep{McQuinn2011} but is not observed in the hydrogen-ionizing background at these redshifts.

Our result emphasizes the importance of understanding the IGM for interpreting measurements of the ionizing background and of reionization, including that of both \HeIIa and \HIa.  In the context of \HeIIa reionization, \citet{DF2009} argued that the rapidly increasing mean optical depth in the \HeIIa \lya line is consistent with ongoing \HeIIa reionization at $z \ga 2.7$. However, they prescribed a relatively slow evolution in the mean free path of ionizing photons. On the other hand, a number of observations show substantial fluctuations in the mean optical depth, even when averaged over large scales \citep{Reimers1997,Zheng2004,Shull2004,Shull2010}. Our model does not address such large-scale fluctuations, because we have not incorporated any spatial information into the calculation.  

This calculation may also have important implications for \HIa reionization, where an apparent rapid increase in the \HIa \lya optical depth has long been attributed to the tail end of reionization \citep{Fan2002, Fan2006}.  \citet{Furlanetto2009b} previously showed that the overlap process of reionization (when ionized bubbles overlap to fill space) does not by itself cause a rapid increase in the ionizing background. We have shown that such an increase can be caused by ``normal" post-reionization processes, through the interaction of a slowly increasing emissivity and slowly decreasing IGM clumping. Whether this occurs during \HIa reionization cannot be said, because it depends sensitively on the evolution of that clumping (which is largely hidden due to the high opacity of the \lya forest beyond $z \sim 6$). However, this \HeIIa analog indicates that a proper interpretation of data regarding \HIa reionization requires careful modelling (and ideally observations) of the IGM and not simply an understanding of the emitting sources.\\

We thank K.~Dixon, G.~Rudie, C.~Steidel, D.~Syphers, and G.~Worseck for helpful conversations, and the anonymous referee for many helpful comments. We also thank G.~Worseck for providing $\Delta z=0.1$ effective optical depth data. This research was partially supported by the David and Lucile Packard Foundation and the Alfred P. Sloan Foundation.

\bibliographystyle{mn2e}

\end{document}